\documentclass[sigconf]{acmart}
\renewcommand\footnotetextcopyrightpermission[1]{}
\usepackage{booktabs} 
\usepackage{enumitem}
\usepackage{balance}
\usepackage[utf8]{inputenc}
\usepackage{multirow}
\usepackage{graphicx}
\usepackage{microtype}
\usepackage{subfigure}
\usepackage{bm}
\usepackage{amsmath}
\usepackage{amsfonts}
\usepackage{framed}
\usepackage{verbatim}
\usepackage{appendix}

\newcounter{myframe}
\newenvironment{numberedframe}[1][]{%
    \refstepcounter{myframe}%
    \begin{framed}%
    \noindent\textbf{Prompt \themyframe: #1}%
    \par\noindent%
    }{%
    \end{framed}%
}
\counterwithin{myframe}{section}

\newcommand{\squishlist}{
 \begin{list}{$\bullet$}
  { \setlength{\itemsep}{0pt}
     \setlength{\parsep}{3pt}
     \setlength{\topsep}{3pt}
     \setlength{\partopsep}{0pt}
     \setlength{\leftmargin}{1.5em}
     \setlength{\labelwidth}{1em}
     \setlength{\labelsep}{0.5em} } }

\newcommand{\squishlisttwo}{
 \begin{list}{$\bullet$}
  { \setlength{\itemsep}{0pt}
     \setlength{\parsep}{0pt}
    \setlength{\topsep}{0pt}
    \setlength{\partopsep}{0pt}
\setlength{\leftmargin}{2em}
\setlength{\labelwidth}{1.5em}
\setlength{\labelsep}{0.5em} } }

\newcommand{\squishend}{
\end{list}  }

\AtBeginDocument{%
  \providecommand\BibTeX{{%
    \normalfont B\kern-0.5em{\scshape i\kern-0.25em b}\kern-0.8em\TeX}}}

\setcopyright{acmcopyright}
\copyrightyear{2018}
\acmYear{2018}
\acmDOI{10.1145/1122445.1122456}

\acmConference[Woodstock '18]{Woodstock '18: ACM Symposium on Neural
  Gaze Detection}{June 03--05, 2018}{Woodstock, NY}
\acmBooktitle{Woodstock '18: ACM Symposium on Neural Gaze Detection,
  June 03--05, 2018, Woodstock, NY}
\acmPrice{15.00}
\acmISBN{978-1-4503-XXXX-X/18/06}

\begin{document}


\title{LLM-Based User Personas for Recommendations at Scale}


\author{Haoting Wang}
\affiliation{%
  \institution{Google Deepmind}
  \city{New York}
  \state{NY}
  \country{USA}
}
\email{haotingwang@google.com}

\author{Haokai Lu}
\affiliation{%
  \institution{Google Deepmind}
  \city{Mountain View}
  \state{CA}
  \country{USA}
}
\email{haokai@google.com}

\author{Zheyun Feng}
\affiliation{%
  \institution{Google}
  \city{Mountain View}
  \state{CA}
  \country{USA}
}
\email{zfeng@google.com}

\author{Jenny Huang}
\affiliation{%
  \institution{Google}
  \city{Mountain View}
  \state{CA}
  \country{USA}
}
\email{huangjenny@google.com}

\author{Yifat Amir}
\affiliation{%
  \institution{Google}
  \city{Mountain View}
  \state{CA}
  \country{USA}
}
\email{yifat@google.com}

\author{Gregory Hinkson}
\affiliation{%
  \institution{Google}
  \city{Mountain View}
  \state{CA}
  \country{USA}
}
\email{gregkhinkson@google.com}

\author{Ben Most}
\affiliation{%
  \institution{Google}
  \city{Mountain View}
  \state{CA}
  \country{USA}
}
\email{bmost@google.com}

\author{Zelong Zhao}
\affiliation{%
  \institution{Google}
  \city{Mountain View}
  \state{CA}
  \country{USA}
}
\email{zelongzhao@google.com}

\author{Kelly (Yixin) Cui}
\affiliation{%
  \institution{Google}
  \city{Mountain View}
  \state{CA}
  \country{USA}
}
\email{kellycui@google.com}

\author{Rein Zhang}
\affiliation{%
  \institution{Google}
  \city{Mountain View}
  \state{CA}
  \country{USA}
}
\email{reinwzhang@google.com}

\author{Fabio Soldo}
\affiliation{%
  \institution{Google}
  \city{Mountain View}
  \state{CA}
  \country{USA}
}
\email{fsoldo@google.com}

\author{Yu Xia}
\authornote{Work performed while at Google.}
\affiliation{%
  \institution{GNucleus AI}
  \city{Mountain View}
  \state{CA}
  \country{USA}
}
\email{peggy.xia@gnucleus.ai}

\author{Nihar Bhupalam}
\affiliation{%
  \institution{Google}
  \city{Mountain View}
  \state{CA}
  \country{USA}
}
\email{niharb@google.com}

\author{Minmin Chen}
\affiliation{%
  \institution{Google Deepmind}
  \city{Mountain View}
  \state{CA}
  \country{USA}
}
\email{minminc@google.com}

\author{Konstantina Christakopoulou}
\affiliation{%
  \institution{Google Deepmind}
  \city{Mountain View}
  \state{CA}
  \country{USA}
}
\email{konchris@google.com}

\author{Lichan Hong}
\affiliation{%
  \institution{Google Deepmind}
  \city{Mountain View}
  \state{CA}
  \country{USA}
}
\email{lichan@google.com}

\author{Ed Chi}
\affiliation{%
  \institution{Google Deepmind}
  \city{Mountain View}
  \state{CA}
  \country{USA}
}
\email{edchi@google.com}

\renewcommand{\shortauthors}{Wang et al.}


\begin{abstract}
Large Language Models (LLMs) offer unprecedented potential for enhancing recommendation systems through their world knowledge and reasoning capabilities. However, existing approaches often rely on structured IDs or offline processing, limiting semantic richness, real-time adaptability, and user-facing interpretability. In this paper, we introduce a novel framework that enables real-time generation of LLM-based user interest personas for a large-scale commercial video recommendation platform. Our method generates natural-language user interest personas that address the exploitation-exploration trade-off by combining the summarization of existing interests with novel topics, directly during serving. To overcome the computational challenges of online LLM inference at a billion-user scale, we design a cost-efficient architecture leveraging knowledge distillation, asynchronous inference, and input optimization via semantically clustered video representations. Extensive offline evaluations, user studies, and live A/B tests demonstrate significant improvements in viewer value. This work bridges the gap between high-level semantic understanding and industrial-scale recommendation, paving the way for more dynamic, explainable, and satisfying personalized experiences.
\end{abstract}

\maketitle
\section{Introduction}

Recently, Large Language Models (LLMs) have begun to reshape the landscape of recommendation systems. Their world knowledge and reasoning capabilities have been leveraged in several ways. Initial applications focused on enhancing item understanding by using LLMs to create richer content embeddings from textual metadata like titles and descriptions \cite{videoannotation2025, episode2025youtube}, and visual information like thumbnails or sampled frames \cite{wang2025serendipitousllm}. Other approaches leverage LLMs by fine-tuning the LLM to generate structured outputs, like content or cluster ids \cite{changpingRAG2025, wang2025serendipitousllm, wang2024llms}, for direct use in traditional retrieval tasks.

Notably, these methods often bypass the generation of natural language, opting for structured IDs to ensure seamless compatibility with existing ID-based recommendation infrastructure. This trade-off sacrifices the rich, nuanced semantics inherent in textual descriptions and precludes the use of these insights in user-facing features, such as human-readable interest summaries.

Moreover, due to significant computational costs and stringent latency constraints, these applications \cite{wang2024llms, wang2025serendipitousllm,changpingRAG2025} operate almost exclusively in an offline or near-offline capacity. This reliance on pre-computation limits the system’s ability to respond to a user’s immediate intent and capture rapidly evolving interests.

In this paper, we propose a novel framework that overcomes these limitations by enabling a real-time LLM-based user understanding service in a large-scale commercial video recommendation platform. Our framework generates high-quality and interpretable user interest personas in natural language during the serving process. Specifically, we make the following two primary contributions:

\begin{itemize}
\item \textbf{High-Quality Natural Language Persona Generation}: We introduce a method to synthesize concise and nuanced textual summaries of user interests from noisy interaction data. Crucially, this approach extends beyond interest summarization; it harnesses the LLM's world knowledge and reasoning capability to infer novel topics for interest exploration, directly mitigating the feedback loop. To make this dual-objective generation scalable, we employ knowledge distillation to transfer complex reasoning capabilities from a large teacher model into an efficient student model.
\item \textbf{Cost-Efficient Online Inference Architecture}: We design and implement an infrastructure capable of supporting online LLM inference at a scale of billions of users, while adhering to strict latency and resource budgets. We achieve this through an asynchronous generation pipeline that decouples inference from live traffic, complemented by model quantization, efficient input structuring, and robust safety mechanisms.
\end{itemize}

We validated the framework through large-scale A/B testing on a large-scale commercial video recommendation platform, which revealed statistically significant improvements in viewer value, evidenced by increased viewer satisfaction and exploration. Furthermore, qualitative findings from user surveys confirmed a high degree of satisfaction with the accuracy of the LLM-generated interest personas.

\section{Related Work}
\noindent \textbf{LLM for User Understanding}
Traditional recommender systems predominantly rely on ID-based sequential models, ranging from recurrent neural network-based \cite{RNNUserUnderstanding} and graph-based approaches \cite{Graphbaseduserunderstanding} to modern Transformer architectures \cite{sasRec, BERT4Rec} to capture user preferences. While efficient, these methods treat items as identifiers, lacking the semantic reasoning required to interpret the underlying motivations behind a user's behavior. 

To address this semantic gap, recent research has pivoted toward LLM-based user understanding. Foundational works like \cite{P5} reformulated recommendation as a sequence-to-sequence language task, demonstrating that LLMs can effectively reason about user behavior given textual prompts. Building on this, approaches like PALR \cite{PALR} demonstrate that fine-tuning LLMs directly on the textual representation of user interaction history allows the model to capture complex user preferences that ID-based methods miss. Similarly, TALLRec \cite{TALLRec} highlights that aligning LLMs with recommendation-specific tasks through fine-tuning significantly enhances their ability to infer user intent from sparse textual data.

However, these implementations exhibit three critical limitations. First, these methods' synchronous design is cost- and latency-prohibitive to deploy at a billion-user scale. Consequently, they are evaluated exclusively on academic benchmarks, which often lack the scale, complexity, and noise characteristic of industrial production environments. Second, these approaches generally fine-tune LLMs to perform traditional recommendation tasks, such as item ranking or binary prediction. As a result, they inadvertently reinforce the biases of the existing system's feedback loop. Lastly, some of these works bypass the generation of explicit, reasoning-based natural language user personas, thereby forfeiting the interpretability and serendipity benefits that semantic user personas offer.

While recent industrial works like Fabbri et al. \cite{spotify_2025} utilize natural-language user profiles, their application is strictly limited to an offline evaluation involving only 47 users. In their framework, the LLM persona acts solely as a passive verifier of existing recommendations rather than a foundational source for generating them. Our work fundamentally advances by embedding LLM personas directly into the live serving path, acting as a real-time retrieval source that actively drives user exploration.

Other industrial frameworks \cite{wang2024llms,wang2024llmsuserexplorationlargescale,wang2025serendipitousllm,Meng_2025} have attempted to introduce reasoning capabilities at scale. However, our framework significantly advances beyond these methods in two critical dimensions: user representation and persona quality. First, restricted by the combinatorial complexity of offline table generation, approaches like \cite{wang2024llmsuserexplorationlargescale,wang2025serendipitousllm,Meng_2025} rely on the cluster labels of only a user's last two watched videos, thereby failing to capture long-term interests. In contrast, our proposed method directly process long contextual watch histories, leveraging structured inputs to successfully reason over both long-term and rapidly evolving interests. Second, prior works constraints the LLM to generate outputs from a small, fixed vocabulary that are restricted to a few hundreds predefined clusters, and is designed strictly for offline batch inference. Conversely, our system acts as a true open-ended reasoning agent, generating free-form natural language interests alongside explicit reasoning traces. These fine-grained user personas capture subtle semantic nuances and provide human-readable explainability directly applicable to user-facing features.

\section{Preliminaries}
\label{sec:preliminary}

\subsection{Hierarchical Planning Paradigm} 
In this paper, we focus on a hybrid hierarchical planning paradigm \cite{wang2024llmsuserexplorationlargescale, wang2025userfeedbackalignmentllmpowered, wang2025serendipitousllm} that leverages LLM to reason about user's interest persona, converting the natural language based interests into item recommendation. Specifically, the recommendation process is decoupled into a high-level language policy and a low-level item policy. During the high-level planning stage, an LLM generates a language policy by predicting a textual description for the user's next interest. At the low level, a classic recommendation model generates a policy that grounds the user interests in item space.

There are many existing approaches to map the textual user interests to the item space. One approach utilizes transformer-based sequential models\cite{vaswani2023attentionneed} paired with restricted nearest neighbor search \cite{scam_2020}. Covington et al. \cite{10.1145/2959100.2959190} retrieves items using nearest neighbor search within the dot-product space of item embeddings and the user embeddings. We build upon this architecture by introducing a semantic constraint: we restrict the nearest neighbor search space to the items that are semantically related to the LLM-generated text interests. This effectively confines the generation space to the user's text interests. 

Another approach employs conventional deep neural network models \cite{bhattacharya2024jointmodelingsearchrecommendations, zamani2018jointmodelingoptimizationsearch, 10.1145/2505515.2505665, 10.1145/3447548.3467127}, which predominantly utilize dual-encoder architectures. By encoding user search queries and past interactions into dense vectors, these models are trained with objectives to maximize the similarity between the user embeddings and relevant item embeddings in a shared space. 

In this framework, we adopt the transformer-based sequential model along with restricted nearest neighbor search. This allows us to effectively take advantage of the already highly optimized sequence models for large-scale item recommendation in production.  
\subsection{User History Clustering}
\label{subsec:user_cluster}

Our framework for generating high-quality user interest personas builds upon existing video clustering methods to structure noisy interaction data. In the context of a large-scale video recommendation platform, we leverage the rich metadata inherent to video content. To identify the optimal user representation for LLM ingestion, we compared two primary approaches:

\noindent \textbf{Embedding-based Clusters} The first approach utilizes a hierarchical video clustering method \cite{pmlr-v139-rajagopalan21a} based on the videos' audio and visual embeddings. This scalable, online, density-based algorithm processes a continuous stream of new videos in real time. This architecture produces a two-level hierarchy of clusters, at the base level are fine-grained clusters, containing items that are all within a fixed distance in one another. These fine-grained clusters are later aggregated into broader macro clusters over time by merging nearby fine-grained clusters. This method is designed to produce stable cluster identifiers.

However, the clusters generated by this method lack semantic meaning. Their organizing principle is perceptual similarity, grouping videos that look and sound alike, rather than a shared abstract concept. This makes them ill-suited for generating the nuanced, human-readable interest labels that are the focus of our work.

\noindent \textbf{Semantic-based Clusters} We adopted the Infinite Concept Personalized Clustering algorithm \cite{christakopoulou2023largelanguagemodelsuser} to group a user's historical watched videos into coherent clusters based on semantic similarity. In this approach, each video is represented by \textit{salient terms}. Salient terms are a set of unigrams and bigrams, each term associated with a weight known as the salience score. Salient term annotations are trained with the video title, uploader name, video description, titles of other videos related to this video, and other text metadata about the video. The salience score is in [0, 1], describing how relevant the salient term is to the video. These terms serve as a high-level semantic representation of the video. The algorithm iterates through a user's interaction history chronologically. For each new item, we compute the cosine similarity between the item's salient terms and the centroids of the existing clusters, then either create a new cluster for this new item, or update the nearest cluster with this new item, based on a similarity threshold. Finally, clusters containing an insufficient number of videos are pruned.

This semantic approach yields highly personalized and interpretable clusters. By organizing content around abstract concepts rather than visual or audio similarity, semantic clustering successfully captures the underlying intent and the topical themes driving user behavior. Indeed, in the offline evaluation study, we found that semantic-based clusters generate higher quality interest personas compared to embedding-based clusters, as shown in Figure \ref{fig:user_representation}(a). Consequently, we utilized semantic-based clustering to structure user histories during the training data collection process.


\section{Method}
This section details our proposed method, focusing on how it addresses two key challenges:  (1) generating high-quality user interest personas; and (2) building a cost-efficient system to support large-scale, per-user inference for a system with billions of users.

\subsection{User Interest Personas Generation} 
We leverage LLMs to perform a complex, dual-function task within a single inference call, creating a holistic user interest persona that balances exploitation and exploration:
\begin{itemize}
    \item \textbf{Summarized Interests}: The first part of the task focuses on exploitation by analyzing a user’s interaction history to synthesize concise, human-readable textual labels that capture their existing preferences.
    \item \textbf{Exploration Interests}: The second part of the task facilitates exploration by leveraging the LLM's vast world knowledge and reasoning capabilities, to generate novel yet semantically related topics for each summarized interest.
\end{itemize}

\subsubsection{User Representation}
To effectively perform the dual-function task of summarization and exploration, the user's interaction history must be represented in a format that is semantically accessible to the LLM. In traditional recommender systems, a user is typically represented by a sequence of item IDs, which are then mapped to dense embeddings. However, such numerical identifiers are opaque to an LLM, as they lack the inherent semantic context required for natural language reasoning. Therefore, our approach necessitates a shift from ID-based to text-based user representation. We studied the best user representation for the summarization task and had the following insights:

\begin{itemize}
    \item \textbf{Insight 1: Video Metadata} We construct a user's persona by leveraging the textual metadata of their watched videos. We found that video titles yielded superior results compared to video descriptions and salient terms. 
    
    We attribute this to the higher signal-to-noise ratio in titles; they provide a concise semantic representation of the content. In contrast, we found that video descriptions often contain significant noise, including promotional text and copyright attributions (e.g., background music credits) irrelevant to the video’s actual topic. Meanwhile, salient terms tend to be overly broad, lacking the specific nuance required for accurate persona generation.
    \item \textbf{Insight 2: Structured Input}  A key finding from our research is that the structure of the input has a profound impact on the granularity of the LLM's output. Feeding the model a simple, unstructured sequence of a user's watched video titles often results in overly broad and abstract interest summaries. We pre-process the user's watch history by grouping videos into semantically coherent clusters before creating the prompt, and found this significantly increased the semantic richness and specificity of LLM summarized user interests.
    \item \textbf{Insight 3: Model Size} We observed that larger models consistently perform better than smaller ones up to a certain scale. 
    
    As shown in Figure ~\ref{fig:user_representation}(c), both the Gemini Pro and Gemini Ultra significantly outperformed Gemini Flash on the summarization task. However, there is a performance plateau when increasing the model size from Pro to Ultra. This indicates a saturation point in the reasoning capabilities required for generating user interest personas.
    
\end{itemize}

The Insight 3 directly motivated our decision to employ knowledge distillation: to harness the power of a large, high-performing model while meeting the cost and latency requirements of an industrial-scale system.

\subsubsection{Distillation}
\label{subsubsec:distillation_data}

To adhere to strict latency and resource constraints in our online serving environment, we designed a unified LLM prompt to perform both interest summarization and exploration tasks within a single inference call. However, the complexity of this dual-objective task necessitates a model with substantial reasoning capabilities. As demonstrated in Sections \ref{subsec:user representation} and \ref{subsec:distillation_eval}, off-the-shelf small models fail to perform these simultaneous tasks effectively.

To address this limitation, we employ knowledge distillation. This approach allows us to transfer the capabilities of large, high-performing models into smaller, efficient architectures, achieving an optimal trade-off between serving cost and generation quality. We utilized Gemini 1.5 Pro ~\cite{team2024gemini}, state-of-the-art model at the time of deployment, as the teacher to generate training data.

\smallskip \noindent \textbf{Data Collection} Our data collection process consists of the following steps:
\begin{enumerate}
    \item \textbf{User Sampling:} We randomly sampled tens of thousands of eligible users and their watch history sequences. Eligibility was restricted to users who consented to data usage for training and possessed a sufficient history of recent, high-satisfaction watch events. We further removed unsafe or sensitive videos from user's watch history.
    \item \textbf{Input Structuring:} To optimize the semantic density of the input, we aggregated the user's recent watch history into semantic clusters using salient term similarity. As demonstrated in our ablation studies (Section 4.1), this clustered format yields superior summarization quality compared to raw sequential inputs. We also removed clusters with few watches to ensure meaningful representation.
    \item \textbf{Multi-step Reasoning Workflow:} To ensure high-quality ground truth labels, we decomposed the unified serving prompt (Prompt~\ref{frame:prompt_2_tasks}) into a multi-step reasoning process as illustrated in Figure ~\ref{fig:training_data}. We employ multiple LLM calls and Chain-of-Thought prompting \cite{cotprompting} to separately generate and refine the summarized interests and exploration topics along with reasoning.
    
\begin{figure}[h!]
    \centering
    \includegraphics[width=0.47\textwidth]{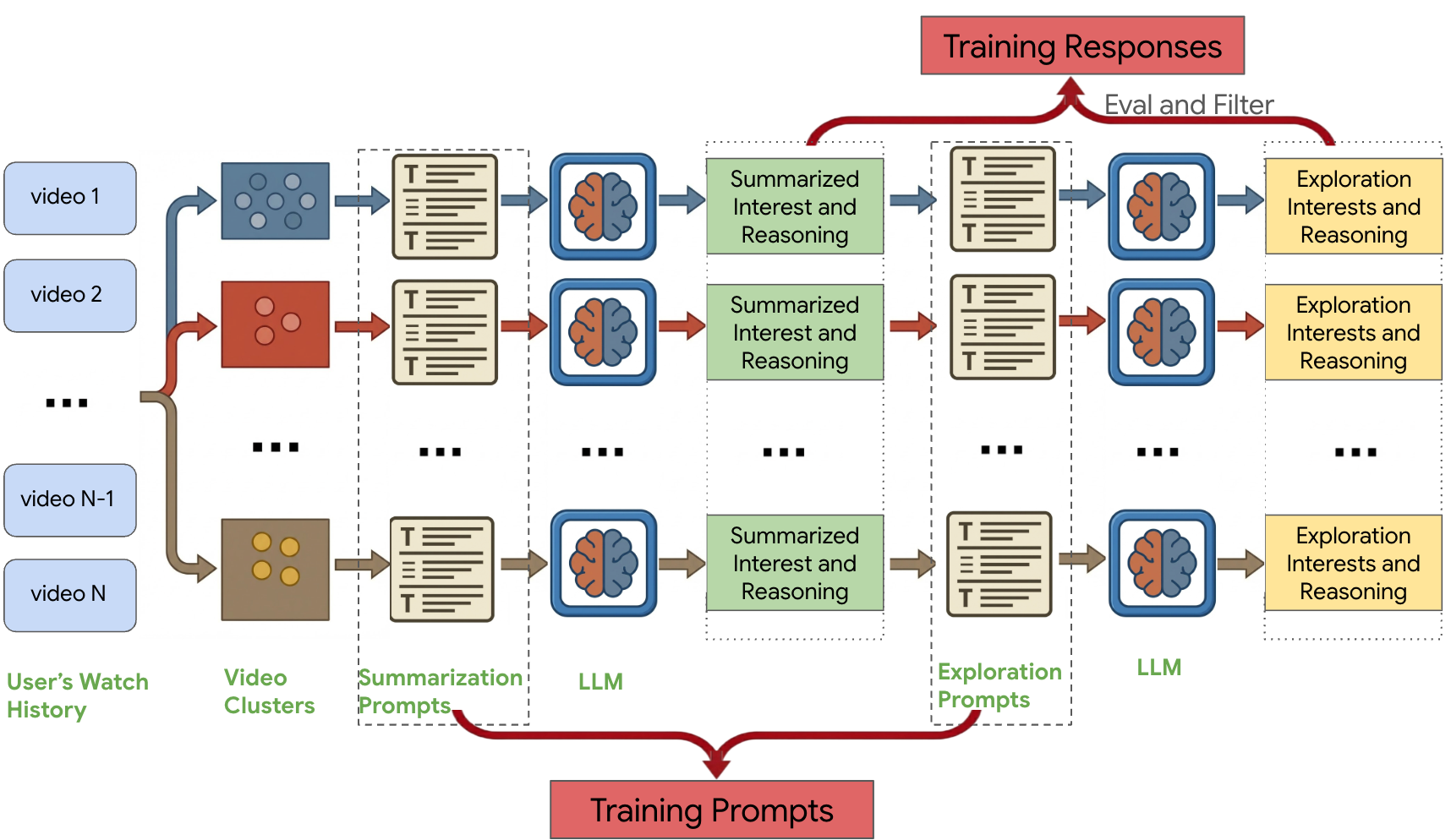}
    \caption{Training Data Collection with Multi-step Reasoning}
    \label{fig:training_data}
\end{figure}

    \item \textbf{Quality Control on Teacher Responses:} We performed a rigorous quality check on the responses: we only kept the responses that contain the same number of interests as the prompt instructed and follow the same format as the prompt instructed.
    \item \textbf{Dataset Splitting:} We take the remaining qualified responses from the previous step, randomly sample 80\% as training data, and rest 20\% as evaluation data. 
\end{enumerate}

This high-quality dataset forms the basis for our distilled models. In the next section, we present the comprehensive online architecture designed to integrate these models into the production environment and manage the full lifecycle of user interest persona.

\subsection{Cost-Efficient Online Serving Architecture}
We designed a robust infrastructure capable of supporting online LLM inference at the scale of billions of users and seamlessly integrating with existing recommendation stacks.

\begin{figure}[h!]
    \centering
    \includegraphics[width=0.43\textwidth]{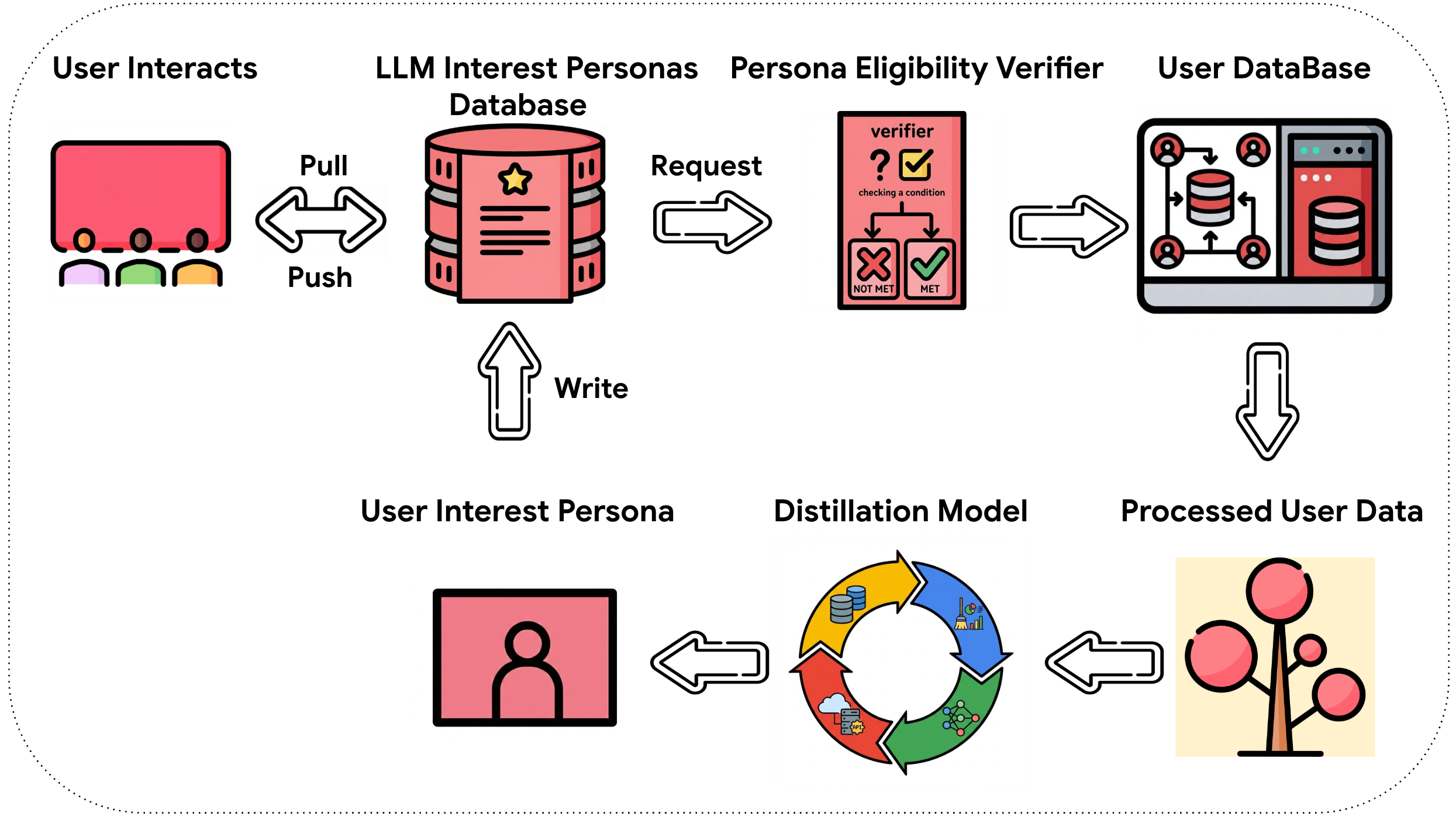}
    \caption{Asynchronous online inference diagram}
    \label{fig:infra}
\end{figure}

\smallskip
\noindent\textbf{Asynchronous LLM Inference} The asynchronous online inference process (illustrated in Figure \ref{fig:infra}) begins when a user visits the website. The system first checks for an existing and up-to-date interest persona in the database. If a valid persona is found, it is immediately retrieved for downstream tasks. If a persona is missing or stale, the system triggers an asynchronous background job to regenerate the persona.  Crucially, to adhere to strict latency constraints, the current request does not wait for this generation. The background job independently retrieves user history, queries the LLM, and updates the database, ensuring the fresh persona is available for the user's subsequent visits. This architecture introduces minimal user-facing latency and optimizes inference costs by decoupling generation frequency from traffic volume. While asynchronous inference inherently involves a trade-off with capturing transient, sudden interests, our system is designed to be highly configurable; both the update frequency and the recency window of the user's watch history can be tuned to mitigate staled updates. Finally, to further save the LLM serving cost, we applied quantization to the student model. 

\smallskip
\noindent\textbf{Preprocess Input Data}
A user is only eligible to have an LLM interest persona if they have sufficient number of qualified watch events. We use a long watch history sequence to ensure we capture the user's long-term and recent interests. We also remove the items that user specifically reported for negative experience to ensure the quality of the interests label. Lastly, we group the filtered user recent watches into few clusters, sample few videos from each cluster, use their titles to represent the cluster, and assemble the sampled titles in the prompt to sample the student model.

\smallskip
\noindent \textbf{Safety} Upon receiving the LLM response, we employ a safety classifier on the LLM output to detect and filter out potentially unsafe or sensitive interest texts, ensuring a safe user experience. In instances where the new response is flagged, the system automatically falls back to the user's previously generated safe and qualified persona.

\smallskip
\noindent\textbf{Interests to Item Retrieval} To enhance personalization, we propose to reuse existing domain-specific retrieval models. We explored the two retrieval options. The first is a Conventional Two-Tower Model \cite{10.1145/3447548.3467127}, where the query tower encodes the LLM-generated persona and the item tower encodes candidate videos, retrieving items by maximizing the cosine similarity between the persona embedding and item embeddings in a shared latent space. The second is a sequential transformer model\cite{vaswani2023attentionneed} that utilizes a constrained nearest neighbor search \cite{scam_2020} between user embeddings and videos semantically related to the LLM generated personas. Those options allow us to connect seamlessly with the existing recommendation system stack. It also enables us to explore the differences between the summarized interests and exploration interests.

\section{Experiments and Results}
\subsection{Offline Study: User Representation }\label{subsec:user representation}
We evaluated different user representation approaches to identify the optimal format for generating high-quality interests summaries.  We conducted a set of offline experiments, measuring the quality of LLM-generated summaries against a ground-truth dataset. The ground truth dataset consists of hundreds of unique users, along with a set of text topics they clicked. These clicks serve as a high-confidence signal of user interests, as selecting a topic directs the user to a page of videos relevant to that specific subject.  We examined user representation across the following dimensions:

\begin{itemize}
    \item Video Representation: title, description, and salient terms. 
    \item Input Structure: Presenting history in chronological order (Prompt~\ref{frame:prompt_sequential_user_summarization}) versus grouping activity into semantically similar or audiovisually similar clusters (Prompt~\ref{frame:prompt_grouped_user_summarization}).
\end{itemize}

\begin{figure}[h]
    \centering
    \includegraphics[width=0.48\textwidth]{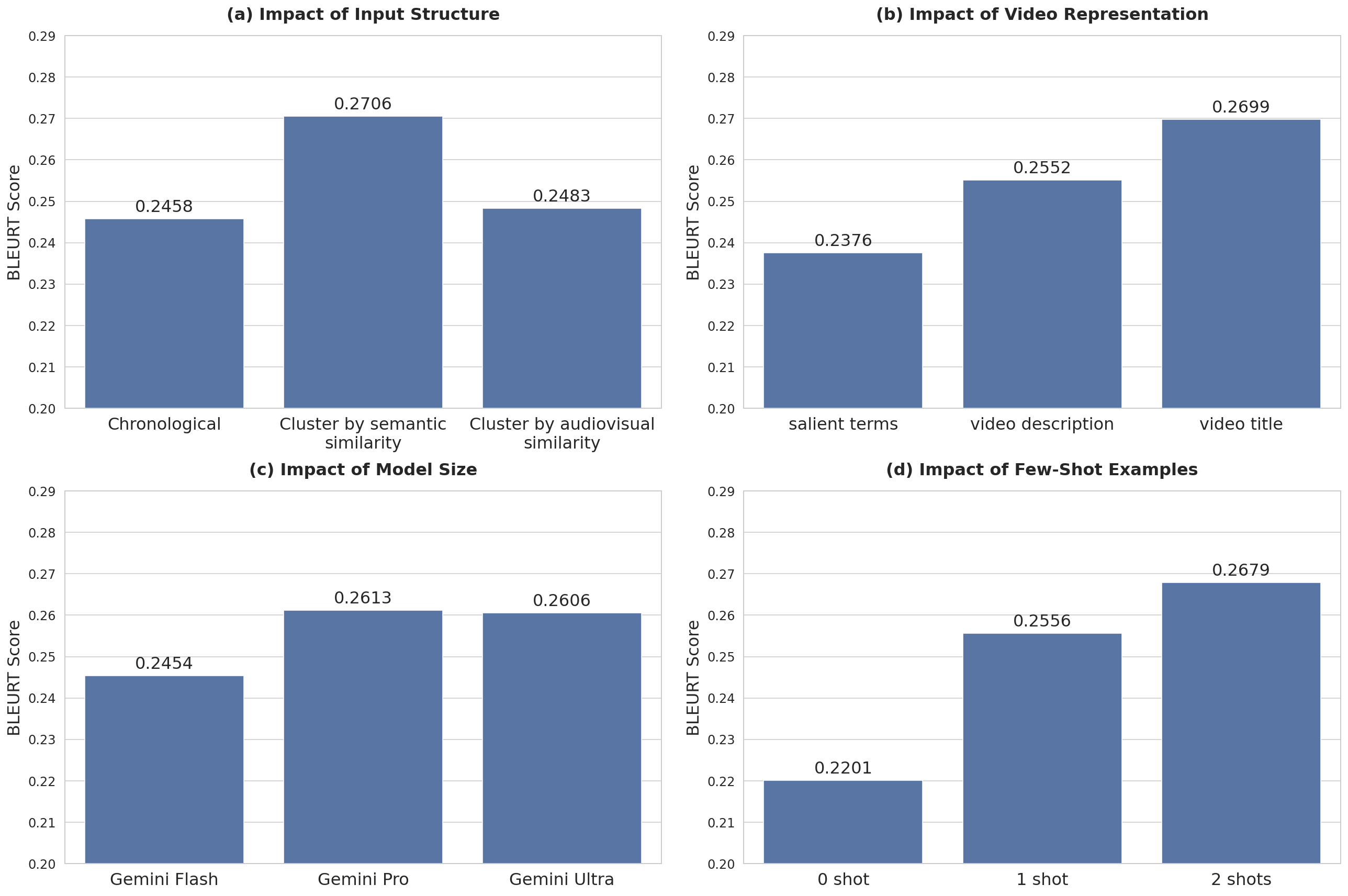}
    \label{fig:grouping_comparison}
    \vspace{-0.15in}
    \caption{Offline evaluation: User representation}
    \label{fig:user_representation}
\end{figure}

We prompted the Gemini models with these varied representations, asked the model to summarize user interests, and evaluated the model's summarized user interests against the ground truth clicked texts using the BLEURT~\cite{sellam2020bleurtlearningrobustmetrics} score. We found that a combination of clustered, title-based inputs and few-shot prompting yielded the highest-quality interest personas, as shown in Figure~\ref{fig:user_representation}(a), (b), and (d). Therefore, for all subsequent experiments, we adopted a clustered, title-based user representation as the standard input for LLMs. 

Additionally, we analyzed the impact of model size on performance. We observed that larger models consistently perform better than smaller ones, as shown in Figure~\ref{fig:user_representation}(c); this finding directly motivated our decision to employ knowledge distillation.

To illustrate the impact of input structure, Table~\ref{tab:qualitative_example} presents a real example comparing the output from the naive sequential representation (Prompt~\ref{frame:prompt_sequential_user_summarization}) against the grouped representation (Prompt~\ref{frame:prompt_grouped_user_summarization}).

\begin{table}[h]
    \centering
    \small
    \begin{tabular}{p{0.35\columnwidth} p{0.58\columnwidth}}
    \toprule
    \textbf{Input Structure} & \textbf{LLM Output} \\
    \midrule
    Sequential \newline (Prompt~\ref{frame:prompt_sequential_user_summarization}) & bengali tv shows \\
    \midrule
    Clustered \newline (Prompt~\ref{frame:prompt_grouped_user_summarization}) & bengali tv series "mon ditey chai" \newline bengali tv series "icche putul" \& "kar kache koi moner kotha" \newline bengali tv series "bojhena se bojhena" \\
    \bottomrule
    \end{tabular}
    \caption{Qualitative comparison of summarized interests}
    \vspace{-3em} 
    \label{tab:qualitative_example}
\end{table}

\subsection{Offline Evaluation: Distillation Performance}
\label{subsec:distillation_eval}
To achieve an optimal balance between performance and serving costs, we employed knowledge distillation. We fine-tuned two student model variants—Gemini Flash and Gemini Nano—using training data described in \ref{subsubsec:distillation_data}. We conducted epoch selection on the evaluation set based on the following metrics:

\begin{itemize}
    \item \textbf{Instruction Following Rate (IFR)}: This metric measures the model's reliability in an industrial setting. A response is considered "instruction-compliant" only if it successfully completes both the summarization and exploration tasks, adheres strictly to the required formatting, and generates the exact number of interest items requested. A high IFR is critical for successful downstream parsing and optimizing TPU resource utilization during serving.
    \item \textbf{BLEURT Score}: To evaluate the quality of the summarization task (exploitation), we utilized the BLEURT score. We treated the summarized interests generated by the teacher model as the reference and compared them against the student model's predictions. This metric assesses how accurately the student model preserves the semantic nuance of the teacher's summaries.
    \item \textbf{Creativity Score}: To evaluate the exploration task, we employed a side-by-side LLM autorater. The autorater compares the pairs of \{summarized interest, exploration interest\} generated by the student model against those from the teacher model, determining which output exhibits greater novelty and creativity. 
\end{itemize}

Table~\ref{tab:distillation_results} details the performance evolution across checkpoints. We observed distinct behaviors across the metrics:
\begin{itemize}
    \item \textbf{Rapid Convergence of Formatting:} The IFR saturated early, indicating that the models quickly learned the structural requirements. The low IFR for all student models at epoch 0 indicates the necessity of distillation if we were ever to serve a small model on live traffic. 
    \item \textbf{Continuous Improvement in Summarization:} BLEURT showed a steady upward trend throughout training. For both student models, the highest BLEURT scores were achieved at epoch 26.20. Notably, the smaller model (Gemini Nano) at epoch 26.20 achieved a score (0.328) comparable to the initial performance of larger variants, justifying the extended training.
    \item \textbf{Creativity Trade-Offs:} Creativity trends varied by model size. Gemini Flash achieved the best balance, continuously improving its novelty scores throughout training. Meanwhile, the smaller model (Gemini Nano) consistently struggled, indicating it likely lacked sufficient parameter capacity to effectively generate novel exploration interests.

\end{itemize}

\begin{table}[htbp]     
   \vspace{-0.5em}
    \centering
    \small 
    \setlength{\tabcolsep}{3.5pt} 
    \renewcommand{\arraystretch}{0.95} 
    \begin{tabular}{l cc cc cc}
        \toprule
        & \multicolumn{2}{c}{\textbf{IFR (\%)}} & \multicolumn{2}{c}{\textbf{BLEURT}} & \multicolumn{2}{c}{\textbf{Creativity}} \\
        \cmidrule(lr){2-3} \cmidrule(lr){4-5} \cmidrule(lr){6-7}
        \textbf{Epoch} & Nano & Flash & Nano & Flash & Nano & Flash \\
        \midrule
        0.00  & 0.07  & 1.82  & N/A & 0.286 & N/A & -0.588 \\
        4.37  & -     & 99.54 & -   & 0.334 & -   & 0.008 \\
        10.92 & 98.97 & 99.75 & 0.321 & 0.338 & -0.012 & 0.016 \\
        26.20 & \textbf{99.08} & \textbf{99.68} & \textbf{0.328} & \textbf{0.345} & -0.018 & \textbf{0.023} \\
        \bottomrule
    \end{tabular}
    \caption{Performance of student models across training epochs}
    \label{tab:distillation_results} 
    \vspace{-0.4em} 
    \raggedright
    \scriptsize \textit{Note: Missing values (-) indicate checkpoints not evaluated. BLEURT and Creativity scores are calculated only on examples that passed instruction following tests. 'N/A' indicates insufficient valid samples to calculate a reliable score due to the near-zero instruction following rate.}
    \vspace{-1.2em} 
\end{table}

\subsection{User Satisfaction On LLM-Generated Interests}

To understand the quality of the LLM summarized interests, we conducted a survey with thousands of users in the United States who had been active on the video recommendation platform during the week preceding the study.

For each participant, we grouped their recent watch history into semantically similar clusters and prompted the LLM to generate an interest for each group. Then we randomly selected one cluster and displayed three representative videos from a cluster. Participants were first asked if they recalled watching these videos. If they answered "Yes," they were presented with the generated interest label and asked to rate it on a 5-point Likert scale (ranging from "Not at all" to "Extremely") based on two questions:
\begin{itemize}
    \item Accuracy: "How closely does [Interest Label] summarize the topic of these videos?"
    \item Preference: "How interested are you in seeing more videos about [Interest Label]?"
\end{itemize}
The survey results indicate a high degree of user satisfaction. Over $80\%$ of respondents reported that the LLM-generated label summarized their watch history "Very Closely" or "Extremely Closely." Furthermore, $71\%$ of users expressed a strong interest in watching more videos related to the generated topic, validating the value of these personas for personalized recommendation.

To rigorously evaluate the LLM personas against existing conventional extractive methods, we conducted a second comparative survey asking users to compare the LLM-generated personas against a control baseline of topics derived from knowledge graph entities\cite{googleknowledgegraph}. The results demonstrated a clear preference for the generative approach: $57\%$ of users strictly preferred the LLM personas, while an additional 
$20\%$ rated them as equivalent to the baseline.

Finally, we analyzed the main reasons for user dissatisfaction. The errors mostly fell into the following four categories: missing some of the user's main interests, inferring interests based on sporadic activity, mentioning outdated interests, and generating repetitive interest labels. Developing mechanisms to continuously refine user personas and address these specific limitations remains an important direction for future work.


\subsection{Live Experiment}

\label{sec:live_exp}
\subsubsection{Experiment Setup}
To examine the proposed method, we conducted live experiments on a commercial video recommendation platform that serves billions of users. We selected an equal amount of non-overlapping user traffic for control and experiment arms, and ran the experiments for over 30 days. The control arm utilized the existing production recommendation stack, without access to the LLM-generated user interest personas.

In the experiment arm, we leveraged video titles as metadata and grouped the user's past watch history using the embedding-based clustering method as described in Section \ref{subsec:user_cluster}. These clusters were used to prompt the student LLM using the unified prompt described in Prompt \ref{frame:prompt_2_tasks}. We utilized the Gemini Nano model at epoch 26.20. While our offline evaluation in Table \ref{tab:distillation_results} indicated that Gemini Flash trained at epoch 26.20 offered higher creativity, the Nano model achieved comparable summarization quality. Ultimately, we prioritized the distilled Gemini Nano model to satisfy the strict latency and cost constraints of our online serving infrastructure.


Upon generating the personas, we randomly sampled one summarized interest and one exploration interest per request. These textual interests were used to guide a sequential transformer model via restricted nearest neighbor search \cite{scam_2020}, effectively mapping the high-level natural language interest personas into specific item candidates. The selected candidates were then fed into the downstream ranking task and presented to the users in an organic way. We deliberately utilized random sampling rather than a deterministic score-based approach to enforce a uniform exploration policy, and ensure the user experiences diverse intra-session discovery rather than repeatedly seeing the same top-scored interests between asynchronous persona updates. Furthermore, this strategy effectively counterbalances the production system's severe recency bias, retrospectively surfacing dormant, long-term interests for more holistic exploration.

\subsubsection{Results and Analysis}

\begin{figure}[h!]
    \centering
    \subfigure[User Watch Time]
    {
    \includegraphics[width=0.45\columnwidth]{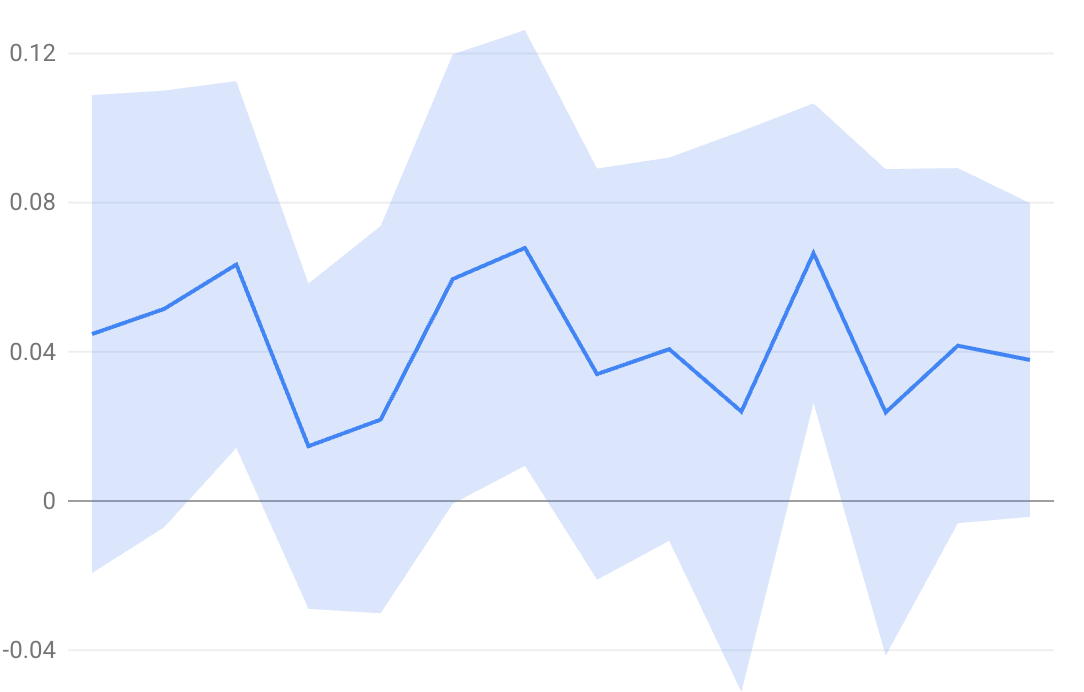}
    }
    \hspace{0.05in}
    \subfigure[\#Active Users]
    {
    \includegraphics[width=0.45\columnwidth]{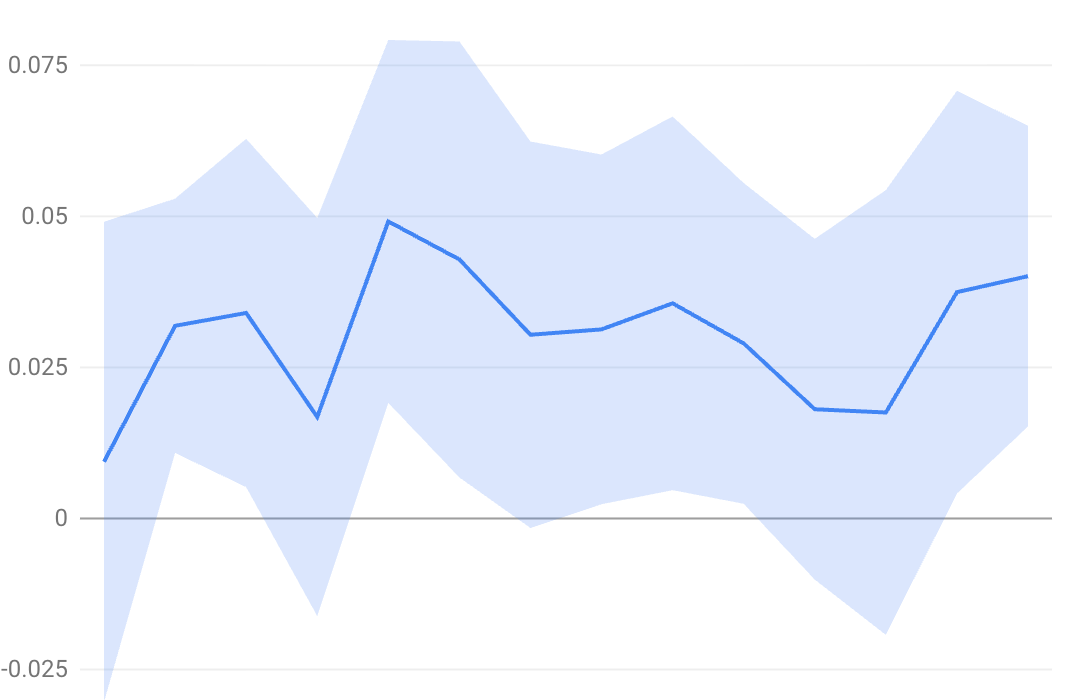}
    }
    \caption{The proposed method drives viewer value. Y-axis shows the degree of improvement of live metrics, X-axis represents the time.}
    \label{fig:engagement}
\end{figure}

\smallskip
\noindent\textbf{Increase in Viewer Value} 
In Figure \ref{fig:engagement}, we measured the impact of integrating the LLM-based user interest personas into the existing recommendation stack. Evaluated against a highly-optimized production baseline serving billions of users, the treatment yielded substantial gains. Figure \ref{fig:engagement}(a) demonstrates a statistically significantly ($p < 0.05$) $+0.04\%$ lift in user's watch time. Similarly, Figure \ref{fig:engagement}(b) shows a statistically significant ($p < 0.05$) $+0.03\%$ increase in number of active users. While these relative percentages may appear numerically small, at our platform's scale, they translate to a massive absolute increase in viewer engagement. This confirms that the proposed natural-language personas provide unique value to the recommender system by introducing nuanced semantic understanding and world knowledge.

To further investigate the impact on viewer value, we segmented our analysis by user activity level. The user population was divided into two cohorts: \textit{casual users}, characterized by lower platform usage, and \textit{core users}, representing the more highly active cohort.  We found that the majority of the improvement in satisfied watch time comes from the casual users. 

We attribute this pronounced effect to two key factors. First, compared to traditional user modeling \cite{Zhao_2023}, the LLM-based interest persona is more adept at inferring user preferences from sparse data. Second, casual users typically exhibit a more concentrated set of interests. Consequently, the interests we sampled from their persona are more likely to achieve comprehensive coverage of their primary affinities, leading to more relevant and satisfying recommendations.

\smallskip
\noindent\textbf{Interest Exploration} 
Our experiments show a significant uplift in users exploring novel topics beyond their established comfort zone. During the live experiment, we observed a statistically significant($p<0.05$) $+0.04\%$ lift in the number of engaged topics, and $+0.03\%$ lift in number of users with multiple lasting engaged topics. These metrics prove a positive behavioral shift: the LLM-generated personas successfully convert serendipitous discovery into sustained, long-term interests.  We further decomposed this impact between the two interest types:

\begin{itemize}
    \item \textbf{Exposure Gap:} Items retrieved via exploration interests received $40.91\%$ fewer impressions than items retrieved via summarized interests, suggesting the downstream ranking models struggle to surface novel content.
    \item \textbf{Recommendation Efficiency:} However, conditional on being presented to the users, items retrieved from the exploration interests are $13.6\%$ more likely to be watched than items retrieved from summarized interests.
\end{itemize}
This discrepancy confirms that the novelty gains are primarily driven by the exploration interest personas, which successfully uncover high-affinity topics that a traditional recommendation system misses.

\smallskip 
\noindent\textbf{Long-Term Impact} 
The impact of LLM-generated user interest personas extends beyond the initial click. We observed that users who watched a recommended video from these personas exhibited prolonged viewer value, resulting in more subsequent visits. This suggests that the language interests in particular have a lasting positive impact, delivering sustained viewer value that encourages deeper and more sustained interaction with the platform.

\section{Conclusion and Future Work}
We introduced a novel LLM-based framework for generating real-time, natural-language user interest personas in a large-scale video recommendation system. Our solution effectively balances summarized and exploratory interests using text-based representations and knowledge distillation to enable cost-efficient online inference. Live experiments and user study confirm significant gains in viewer value, especially for casual users. This work demonstrates that LLMs can be successfully integrated into the core serving path of industrial recommendation systems, not just as offline data processors, but as real-time reasoning engines that drive measurable viewer value.

Future work includes extending user personas to multimodal and cross-domain data, collecting user feedback and enabling user-in-the-loop personas refinement, and developing more interactive use cases of the semantic user personas in the recommendation system.

\begin{appendix}
\appendixpage
\section{Prompts for User Interest Summarization}
This section contains the detailed prompts used to generate summarized user interest personas with the LLM. 

\medskip \noindent Below is the two-shot prompt for the user interests summarization task, where the user activity is represented in a sequential, chronological order

\begin{numberedframe}
\label{frame:prompt_sequential_user_summarization}
\noindent I'm a brilliant video topic summarization expert that speaks all languages. Given a set of videos a person watched, I can describe the interests of that person and explain why respectively. I can also wrap interests of that person using **.

\bigskip \noindent For example:
A person watched videos with titles:  $video\_metadata_1$,  ...  $video\_metadata_{m}$ 

\noindent My output is:

\noindent [Group 0]: **<Summarized Interests 0>**: <Reasoning 0>

\noindent [Group 1]: **<Summarized Interests 1>**: <Reasoning 1>

\noindent [Group 2]: **<Summarized Interests 2>**: <Reasoning 2>

\noindent [Group 3]: **<Summarized Interests 3>**: <Reasoning 3>

\bigskip\noindent As another example:
A person watched videos with titles:  $video\_metadata_1$,  ...  $video\_metadata_{m}$ 

\noindent My output is:

\noindent [Group 0]: **<Summarized Interests 0>**: <Reasoning 0>

\noindent [Group 1]: **<Summarized Interests 1>**: <Reasoning 1>

\noindent [Group 2]: **<Summarized Interests 2>**: <Reasoning 2>

\noindent [Group 3]: **<Summarized Interests 3>**: <Reasoning 3>

\bigskip \noindent Now, if a person watched videos with $video\_metadata_1$,  ...  $video\_metadata_{m}$, My output is:

\end{numberedframe}

\medskip \noindent Below is the two-shot prompt for the user interests summarization task, where the user activity is first grouped into different clusters, then represented by group.

\begin{numberedframe}
\label{frame:prompt_grouped_user_summarization}
\noindent {I'm a brilliant video topic summarization expert who speaks all languages. Given a few groups of videos a person watched, I can describe the interests of that person for each group and explain why, respectively. I can also wrap the interests of that person using **.}

\bigskip\noindent For example:
A person watched the following groups of videos:

\noindent [Group 0]: $video\_metadata_1$,  ...  $video\_metadata_{m_0}$ 

\noindent [Group 1]: $video\_metadata_1$,  ...  $video\_metadata_{m_1}$ 

\noindent [Group 2]: $video\_metadata_1$,  ...  $video\_metadata_{m_2}$ 

\noindent[Group 3]: $video\_metadata_1$,  ...  $video\_metadata_{m_3}$ 

\noindent My output is:

\noindent [Group 0]: **<Summarized Interests 0>**: <Reasoning 0>

\noindent [Group 1]: **<Summarized Interests 1>**: <Reasoning 1>

\noindent [Group 2]: **<Summarized Interests 2>**: <Reasoning 2>

\noindent [Group 3]: **<Summarized Interests 3>**: <Reasoning 3>

\bigskip\noindent As another example:

\noindent A person watched the following groups of videos:

\noindent{[Group 0]}: $video\_metadata_{01}$,  ...  $video\_metadata_{0m_0}$ 

\noindent{[Group 1]}: $video\_metadata_{11}$,  ...  $video\_metadata_{1m_1}$ 

\noindent{[Group 2]}: $video\_metadata_{21}$,  ...  $video\_metadata_{2m_2}$

\noindent  My output is:

\noindent [Group 0]: **<Summarized Interests 0>**: <Reasoning 0>

\noindent [Group 1]: **<Summarized Interests 1>**: <Reasoning 1>

\noindent [Group 2]: **<Summarized Interests 2>**: <Reasoning 2>

\bigskip\noindent Now, a person watched following <Num\_Groups> groups of videos:

\noindent{[Group 0]}: $video\_metadata_{01}$,  ...  $video\_metadata_{0m_0}$ 

\noindent{[Group 1]}: $video\_metadata_{11}$,  ...  $video\_metadata_{1m_1}$ 

\noindent{[Group 2]}: $video\_metadata_{21}$,  ...  $video\_metadata_{2m_2}$ 

\noindent{[Group 3]}: $video\_metadata_{31}$,  ...  $video\_metadata_{3m_3}$ 

\noindent My output is:
\end{numberedframe}

\section{Prompts for Interest Summarization and Exploration}

\begin{numberedframe}
\label{frame:prompt_2_tasks}
\noindent I'm a brilliant video topic summarization expert who speaks all languages. Given a few groups of videos a person watched, I can complete the following 2 tasks:

Task 1: Describe the interests of that person for each group and explain why, respectively. I can wrap the interests of that person using **.

Task 2: For each summarized interest in Task 1, think of 3 creative and exploratory interests that are relevant to that summarized interest, but also novel and provide new perspectives. I can wrap the exploration interests in this task using \&\&.

\bigskip\noindent For example:
A person watched the following groups of videos:

\noindent [Group 0]: $video\_metadata_1$,  ...  $video\_metadata_{m_0}$ 

\noindent [Group 1]: $video\_metadata_1$,  ...  $video\_metadata_{m_1}$ 

\noindent [Group 2]: $video\_metadata_1$,  ...  $video\_metadata_{m_2}$ 

\noindent[Group 3]: $video\_metadata_1$,  ...  $video\_metadata_{m_3}$

\noindent My output is:

\noindent [Group 0]: 

Task 1:
**<Summarized Interests 0>**: <Reasoning 0>

Task 2:
\&\&<Exploration interests 0\_0>\&\&, \&\&<Exploration interests 0\_1>\&\&, \&\&<Exploration interests 0\_2>\&\&

\noindent [Group 1]: 

Task 1:
**<Summarized Interests 1>**: <Reasoning 1>

Task 2:
\&\&<Exploration interests 1\_0>\&\&, \&\&<Exploration interests 1\_1>\&\&, \&\&<Exploration interests 1\_2>\&\&

\noindent [Group 2]: 

Task 1:
**<Summarized Interests 2>**: <Reasoning 2>

Task 2:
\&\&<Exploration interests 2\_0>\&\&, \&\&<Exploration interests 2\_1>\&\&, \&\&<Exploration interests 2\_2>\&\&

\noindent [Group 3]: 

Task 1:
**<Summarized Interests 3>**: <Reasoning 3>

Task 2:
\&\&<Exploration interests 3\_0>\&\&, \&\&<Exploration interests 3\_1>\&\&, \&\&<Exploration interests 3\_2>\&\&

\noindent As another example ...

\noindent Now, a person watched following <Num\_Groups> groups of videos: 
...

\noindent My output is:

\end{numberedframe}

\end{appendix}

\clearpage
\balance
\bibliographystyle{ACM-Reference-Format}
\bibliography{sample_bib}

\end{document}